# Cyclic $C_4^+$ as the carrier of the diffuse interstellar band at 503.9 nm?

Serge. A. Krasnokutski[1][2], Lisa Ganner[3], Milan Ončák[3], Florian Foitzik[3], Stefan Bergmeister[3], Fabio Zappa[3], Paul Scheier[3], Elisabeth Gruber[3].

Abstract

The diffuse interstellar bands (DIBs) have remained a mystery in astronomy since their discovery over a century ago. The only currently known carrier is $C_{60}^+$ responsible for five DIBs, while more than 550 are yet to be interpreted. The spectra of short carbon chain cations ($C_n^+$), which are considered one of the most promising classes of species for the role of carriers of DIBs, are successfully recorded using He-tagging spectroscopy. The comparison of laboratory spectra with the observations demonstrates a close match of two absorption bands of $C_4^+$ with the broad DIB at 503.9 nm. This defines a high abundance of these ions in the interstellar medium (ISM), which should exceed those of other similar-sized carbon chain cations. It is anticipated that all other short carbon chain cations will exhibit linear geometry and, as a consequence, will have a long vibrational progression. However, the distinctive cyclic geometry of $C_4^+$ is postulated to underpin the elevated abundance of these ions in the ISM, as well as the distinctive spectrum of this ion, which displays a single strong, relatively narrow absorption band that exceeds in intensity all other absorption bands in the visible range.

[1] Corresponding author sergiy.krasnokutskiy@uni-jena.de

[2] Laboratory Astrophysics Group of the Max Planck Institute for Astronomy at the Friedrich Schiller University Jena, Helmholtzweg 3, D-07743 Jena, Germany

[3] Institut für Ionenphysik und Angewandte Physik, Universität Innsbruck, Technikerstr. 25, 6020 Innsbruck, Austria

1. Introduction

The diffuse interstellar bands (DIBs), which are absorption bands of the dense parts of the interstellar medium (ISM), were discovered more than a hundred years ago (Heger 1922) and remain mainly unassigned. Only relatively recently five DIBs were assigned to $C_{60}^+$ (Campbell et al. 2015; Cordiner et al. 2019). The missing polarization and observed profiles of the DIBs resulted in a general conclusion that the carriers should be carbonaceous molecules or ions present in the gas phase(Cox et al. 2017; Cox et al. 2011; Geballe 2016; Snow 2001).

Absorption bands of $C_3$ and $C_2$ molecules were detected along many different lines of sight (LoS) (Hinkle, Keady, & Bernath 1988; Lambert, Sheffer, & Federman 1995; Maier et al. 2001; Sonnentrucker et al. 2018). The carbon chains can also very efficiently grow even at low temperatures of the molecular clouds in the ISM due to C or $C^+$ addition (Taniguchi, Gorai, & Tan 2024). Therefore, carbon chains $C_n$ or carbon chain derivatives with different atoms at the ends like $HC_n$, $HC_nH$, $OC_n$, etc. were considered to be very promising candidates for carriers of DIBs (Douglas 1977). Multiple studies were conducted on these molecules (Jochnowitz & Maier 2008; Nagarajan & Maier 2010; Rademacher, Reedy, & Campbell 2022; Reedy et al. 2022; Taniguchi, et al. 2024). Although some matches with observations were found, such as the famous case of the $H_2C_3$ molecule (Oka & McCall 2011), none of the molecules were ultimately confirmed as carriers of DIBs.

Surprisingly, many spectra of the promising species, namely small $C_n^+$, were not recorded with a quality that allows direct comparison to observational spectra. Only relatively recently, the spectra of $C_5^+$ and $C_6^+$ were recorded (Colley, Orr, & Duncan 2022; Reedy, et al. 2022). However, no match with the observed DIBs was reported. Moreover, the observed spectra demonstrated long vibronic progressions for both measured electronic transitions making these species bad candidates for the carriers of DIBs, whose intensities are usually barely correlated with the intensities of other DIBs along different LoS (Smith et al. 2021). In the present communication, we present measurements of small clusters of carbon cations (with $n$ = 3-6). We used the same He-tagging spectroscopy technique that was used for the identification of $C_{60}^+$ as carriers of DIBs (Campbell, et al. 2015). Unlike those earlier experiments, where He-tagged ion species were produced in a cryogenic multipole trap via helium buffer gas cooling, we utilized a recently developed method that employs multiply charged helium nanodroplets to efficiently form helium-tagged molecular ions. This method offers the advantage of controlling the number of attached helium atoms, ranging from a single helium atom to several dozen.

2. Results and discussion

The mass-selective He-tagged spectra of $C_n^+$ (n = 3-6) are shown in Figure 1. For these measurements, ions with a mass-to-charge ratio greater than 92 were selected, including $C_3^+He_m$ (14 ≤ m ≲ 141), $C_4^+He_m$ (11 ≤ m ≲ 138), $C_5^+He_m$ (8 ≤ m ≲ 135), and $C_6^+He_m$ (6 ≤ m ≲ 132). These ions were irradiated with the laser beam, and bare $C_3^+$, $C_4^+$, $C_5^+$, and $C_6^+$ ions were detected as a function of the laser wavelength.

The measured spectra are in perfect agreement with the previous measurements obtained by He-tagging spectroscopy for $C_5^+He$ (Reedy, et al. 2022). The main difference, due to the larger number of He atoms bound in this mode, is the observed blue shift of about 1.58 nm. They are also in general agreement with mass selective spectra of $C_6^+$ (Campbell & Dunk 2019; Colley, et al. 2022). However, our measurement is quite different compared to the matrix-isolation spectra reported for the $C_6^+$ (Fulara et al. 2004). This shows the difficulties with the attribution of the spectral absorption features of different clusters isolated in inert matrices. For $C_3^+$ and $C_4^+$ there is no literature available to compare.

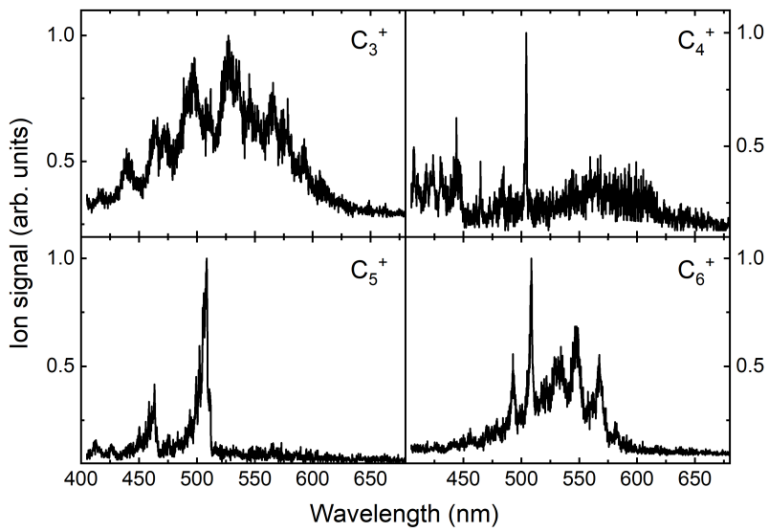

Figure 1. Survey He-tagging spectra of $C_n^+$ clusters. For each frame, the ion signals are normalized to the maximum value.

The spectra of $C_3^+$, $C_5^+$, and $C_6^+$ are similar showing the long vibronic progressions, which are attributed to the considerable change in the length of the C=C bonds in the linear chains after electronic transitions. Therefore, these ions are also not good candidates for the carriers of DIBs. However, the spectrum of $C_4^+$ is distinctly different showing a single strong absorption band near 504 nm, which also points out to a different structure of this ion. The broad continuous absorption in the 450-400 nm range is mainly caused by the amplification of the noise during the laser power correction procedure. In addition, there are a few weak narrow absorption bands on top. Additional scans in the 300–400 nm range did not reveal any notable absorption in the UV range.

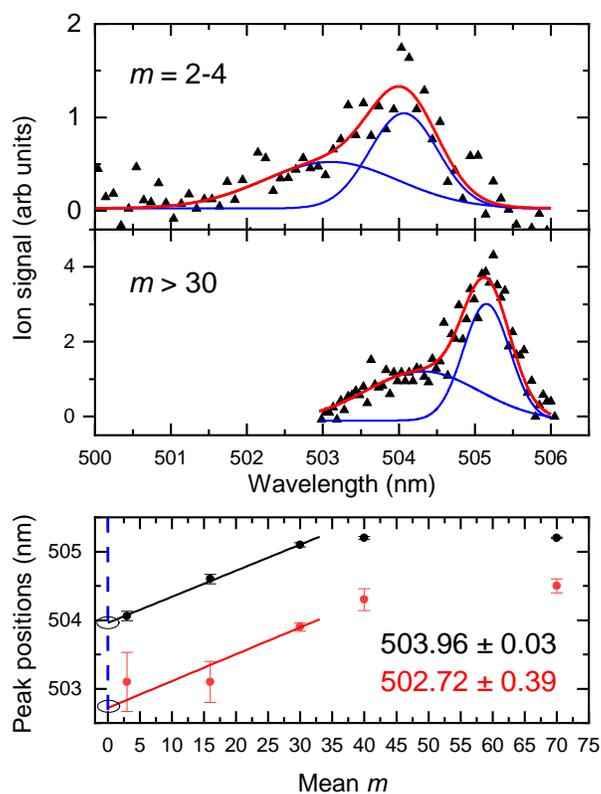

Figure 2. The evaluation of the position of the absorption bands of $C_4^+$ in the gas phase. The two upper images show two measured absorption spectra of $C_4^+He_m$ and the lower one shows the analyzed positions of the $C_4^+He_m$ absorption bands as a function of the average number of He taggants $m$. The individual data points in the lower frame are obtained by fitting the experimental spectra with two Gaussian profiles, as shown in the upper frames. Assuming a linear shift for small $m$, we determine the position of the absorption bands for $m = 0$.

A DIB with a similar band position and width was detected along multiple LoS. It was first reported at 503.91 nm with a full width at half maximum (FWHM) = 1.79 nm as "uncertain or questionable" (Jenniskens & Désert 1994). However, later a relatively strong DIB at 503.86 nm with FWHM = 2.69 nm was reported without mentioning an uncertainty of the detection (Tuairisg et al. 2000). In the most recent DIB catalog, the band at 503.9 nm with FWHM = 1.79 nm is again listed as a possible candidate as it blends with stellar and telluric lines.(Fan et al. 2019; Sonnentrucker, et al. 2018) In all studies, the intensity of this DIB is found to correlate with the optical extinction of clouds, which supports the real nature of this DIB.

To better compare our laboratory data with the observational spectra, we measured the positions of the $C_4^+He_m$ absorption bands as a function of the average number of He taggants $m$. The results are shown in Figure 2. For small $m$, the shift caused by the attachment of He atoms exhibits for several cations a linear trend (Kaiser et al. 2018; Kappe et al. 2023; Meyer et al. 2021; Reedy, et al. 2022). Whereas for m > 30, the addition of further He atoms gives a very small or virtually no shift resulting in a narrower bandwidth of measured absorbtion bands. Therefore, this procedure allows us to obtain the shift caused by addition of a He atom (0.45 Å) and position of the main absorption band of $C_4^+$ for $m$ = 0 (503.96 nm) with a precision comparable to those obtained for astronomical observations.

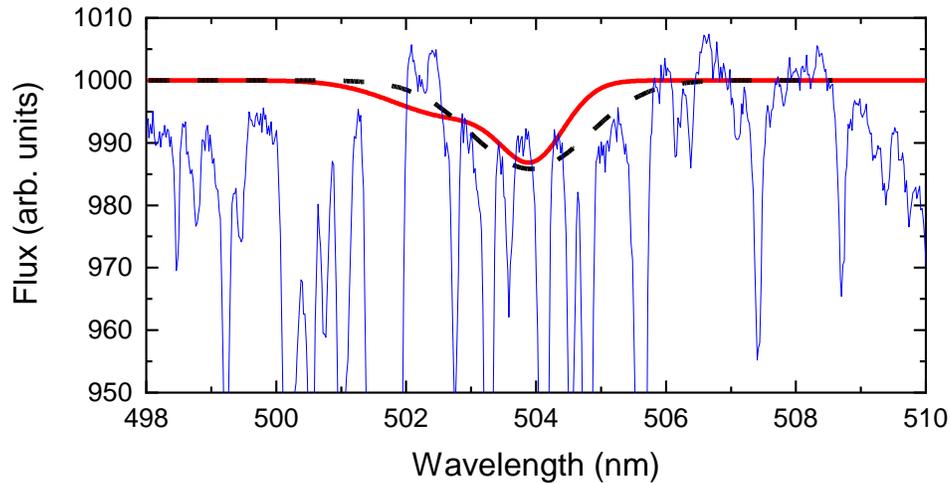

Figure 3. Comparison of experimental and observational spectra from ISO database. The solid red line shows the attenuation of flux due to the absorption of $C_4^+$. The solid blue line is the observational spectrum toward HD 183143 and the dashed black line is the DIB evaluated in previous studies (Jenniskens & Désert 1994).

Figure 3 shows the comparison of the laboratory $C_4^+$ absorption line and the observed DIB. In addition, we are showing the observational spectrum obtained from the ESO Science Archive Facility (HD183143_ADP.2014-05-15T16_14_28.937) to demonstrate the quality of observational data. The estimated photon flux reduction due to the absorption of $C_4^+$ matches fairly well with the DIB reported in previous studies, which is shown by the dashed line in the figure (Jenniskens & Désert 1994; Sonnentrucker, et al. 2018). The bandwidth is considerably higher than anticipated for the extent of the rotational envelope. Consequently, it is likely that the observed broadening is due to lifetime effects, and that the band profile is not significantly influenced by the temperature of the species. However, it is still possible to anticipate a small increase in the bandwidth due to the higher temperatures of the species in the ISM compared to the temperature of He-tagged ions used in the experiment. This would result even in a better match with the evaluated DIB. However, the primary source of uncertainty is the observational spectra. The evaluation of the profile and intensity of this DIB is largely dependent on the accurate correction of the baseline and subtracting all telluric and stellar bands. The large uncertainty of astronomical observations can be demonstrated by different evaluations of the FWHM of this DIB given in different studies that vary by more than 1.5 times (2.69 nm and 1.79 nm) (Sonnentrucker, et al. 2018; Tuairisg, et al. 2000). Considering that $C_4^+$ should be present in the ISM and the close match of our experimental spectrum with the estimated DIB, we conclude that the absorption of $C_4^+$ is a very likely cause of this DIB.

The exact geometry of $C_4^+$ is difficult to define. Different isomers are found to be most stable on different levels of theory. In a comprehensive computational study, more than 30 isomers were found for $C_4^+$, $C_4^-$, and $C_4$ (Wang & Withey 2018). The most energetically favorable isomers, which are very close in energy, have cyclic and linear forms, while other forms are much higher in energy. Therefore, we did not consider them in our analysis. In earlier studies, the linear structure was found to be the most energy-favorable configuration for the neutral and anionic species, whereas the cyclic structure is the most energy-favorable configuration for the cationic species (Wang & Withey 2018). In contrast, our computations at the CCSD/aug-cc-pVTZ level of theory predict that the linear structure is the most energetically favorable for cations as could be seen in Figure 4. In experiments, a linear geometry of neutral $C_4$ was found using infrared laser spectroscopy (Moazzenahmadi, Thong, & Mckellar 1994), while the cyclic $C_4$ was detected in the Coulomb explosion imaging after electron photodetachment from $C_4^-$ (Zajfman et al. 1992). Despite the linear structure of $C_4^+$ being predicted to be the most energetically favorable by our calculations, the UEOMCCSD-FC/aug-cc-pVTZ method indicates that it exhibits significant alterations in the C=C bond lengths upon transition in the excited states. Therefore, in this respect, it is highly similar to other $C_n^+$ chains. This is also confirmed by MRCI(9,9)/aug-cc-pVTZ calculations as can be seen in Figure A2. The predicted large geometry change upon electronic transitions results in a vibronic spectrum with a long vibrational series as can be seen in Figure A3. Therefore, the linear $C_4^+$ would have a spectrum with many absorption lines similar to other measured $C_n^+$ ions, which were demonstrated to have linear geometry (Reedy, et al. 2022; Van Orden & Saykally 1998; von Helden et al. 1993). This does not match the experimental spectrum and allows us to exclude the linear structure from consideration. Therefore, $C_4^+$ should have cyclic geometry, which was also found to be most energetically favorable at different levels of theories (Wang & Withey 2018).

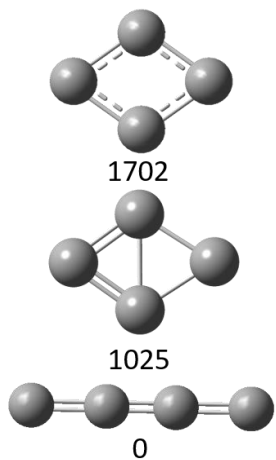

Figure 4. Molecular structures of the $C_4^+$ obtained at CCSD/aug-cc-pVTZ level of theory. The numbers provide the relative energies in Kelvin.

As illustrated in Figure 4, the energy difference between symmetric and asymmetric rhombic cyclic structures is notably minimal. Nevertheless, the computations indicate that the symmetric rhombic structure exhibits electronic transitions in the near-infrared (NIR) range that do not align with the experimental observations (see Figure A4 and Table A1). More precise calculations of the vibronic spectra of cyclic $C_4^+$ are unfortunately complicated due to strong wave function mixing found for this ion (Hochlaf, Nicolas, & Poisson 2007).

The unambiguous identification of $C_4^+$ as a carrier of DIBs would require the detection of other DIBs associated with its absorption bands. However, other absorption bands of $C_4^+$ are considerably weaker, as can be seen in Figure 1. They are mainly concentrated in the 400-450 nm range and are at least twice as faint. In addition, most of them are much broader and have a FWHM of around 3 nm. The intensity of these bands is further reduced when the broad continuous absorption, which is mainly amplified noise, is subtracted. There are no reported DIBs with close parameters. However, the detection of these weak blue bands in astronomical spectra is challenging due to a considerably lower quality of observational spectra in this wavelength range and the general difficulty in detecting broad absorption bands. Therefore, detecting these bands would require observational spectra of higher quality with proper baseline correction having a focus on identifying the broad and weak DIBs. Furthermore, we cannot completely exclude the possibility that other absorption bands belong to different isomers of $C_4^+$, which may have a lower abundance in space.

## 3. Astrophysical Implications

We estimated the column density of $C_4^+$ along LoS toward HD183143 to be around $1.7 \times 10^{14}$ cm$^{-2}$. To do this, we used the calculated oscillator strength f = 0.01 for the electronic transition $D_2(0) \leftarrow D_0(0)$ of cyclic $C_4^+$ and the evaluated equivalent width $W_\lambda$ = 0.38 Å reported for this DIB (Jenniskens & Désert 1994). It should be noted that both values are subject to a large uncertainty. This DIB was reported several times as possible, which means that $W_\lambda$ can be much smaller. In addition, the calculated oscillator strength value is quite imprecise. Although the column density of $C_4^+$ may be somewhat lower, it should still be significant so that its absorption is detected. For comparison, the estimated column density of $C_{60}^+$ along the same LoS is about $2 \times 10^{13}$ cm$^{-2}$ (Campbell et al. 2016). This results in about an equal amount of carbon being locked in $C_4^+$ and $C_{60}^+$ species. The relatively high abundance of $C_4^+$ can be rationalized considering that this small molecule can be efficiently formed via bottom-up chemistry by C atom or C cation addition (Taniguchi, et al. 2024). Additionally, it should be more chemically stable in the ISM compared to other $C_n^+$ clusters. This is due to the same reasons that make its spectrum look special. It is anticipated that all other small $C_n^+$ clusters at least with *n* < 7 will exhibit a linear geometry, rendering them highly reactive even at low temperatures of molecular clouds in the ISM (Van Orden & Saykally 1998). This reactivity ultimately leads to their efficient destruction due to chemical reactions. In contrast, the $C_4^+$ ion, which was predicted to have a cyclic geometry, exhibits strikingly different behavior (Hochlaf, et al. 2007; Wang & Withey 2018). The cyclic structures are much more chemically inert compared to the linear isomers of other C$_n$ ions. This is because the terminal carbon atoms in the linear geometries have unsaturated bonds and are highly reactive. Therefore, there is a less efficient chemical destruction of $C_4^+$ compared to other short $C_n^+$ chains, resulting in a higher abundance of this ion. Additionally, electronic transitions in linear chains result in a notable change in the C-C bond lengths, which in turn produces a long series of vibronic bands. Therefore, the detection of absorption bands of linear chains will be complicated even under conditions of a similar abundance.

$C_4^+$ should still be less stable compared to $C_{60}^+$. However, a much more efficient formation pathway through the $C_3 + C^+ \rightarrow C_4^+ + h\nu$, $C_3H + C^+ \rightarrow C_4^+ + H$, $C_3^+ + C \rightarrow C_4^+ + h\nu$, $C_3H^+ + C \rightarrow C_4^+ + H$ reactions overcompensate for the destruction. The upper limit for the column density of the neutral linear C$_4$ toward ζ Ophiuchi was estimated to be around $10^{13}$ cm$^{-2}$ (Maier, Walker, & Bohlender 2002). Additionally, a very low abundance of C$_2$ and C$_3$ molecules was found toward HD183143 (Oka et al. 2003). This assumes a high ionization fraction of the C$_n$ chains along these LoS, consistent with the high ionization fraction found for the C$_{60}$ molecules, for which no absorption bands of neutral molecules were detected (Rouillé, Krasnokutski, & Carpentier 2021).

The finding of $C_4^+$ possibly being the carrier of the DIBs shows the importance of the chemical stability of the species for achieving a higher abundance in the ISM. Other short $C_n^+$ species are not optimal candidates for being DIB carriers, as their abundances are anticipated to be lower due to their higher chemical reactivity. Additionally, the oscillator strength of the electronic transitions is distributed among multiple vibronic bands, reducing their intensities. However, when *n* becomes sufficiently large, another type of electronic transition, which is approximately 100 times more intense, results in absorption in the visible range. The detection of such huge absorption bands of long carbon chains could still be possible even with a much lower abundance of these species. Therefore, long carbon chains could be good candidates for the DIB carriers if these electronic transitions also result in weak vibronic bands making the detection of these species more straightforward. The spectra of cyclic $C_{2n}^+$ (*n* = 6-14) species were recently recorded and compared with the known DIBs but no convincing match was found (Buntine et al. 2021; Rademacher, et al. 2022). However, these cyclic chains were produced at high temperatures that favor the formation of the lowest energy structures. If we assume the bottom-up formation of carbon chains at low temperatures of the ISM by C atom addition, linear geometry should be expected. This is because the transition from linear to cyclic geometry has a quite high-energy barrier. Short $C_n$ chains with $n < 10$ are linear (Van Orden & Saykally 1998; von Helden, et al. 1993; von Helden et al. 1991), which defines the initial geometry. At low temperatures of the ISM, overcoming the substantial energy barrier needed to elongate the linear chain and close the ring may not be readily achievable. Consequently, long linear $C_n^{0/+}$ chains may still be present in the ISM and may therefore be considered promising candidates for carriers of the DIBs. Additionally, for the $C_n^+$ species with $n$ = 7-9 both cyclic and linear isomers were detected in the experiments (Fulara et al. 2005; Van Orden & Saykally 1998). Although it is commonly considered that the most energy-favorable isomers should dominate in space, the other isomers may still have a higher abundance due to higher stability or faster formation rate. Therefore, cyclic isomers of other short $C_n^+$ could also be promising candidates for the DIB carriers.

## 4. Acknowledgments

The authors are grateful to FWF (Grant-DOI 10.55776/I6221, Grant-DOI 10.55776/V1035, Grant-DOI 10.55776/W1259, and Grant-DOI 10.55776/P34563). SAK is grateful to DFG (grant Nos. 413610339). The computational results presented have been partially achieved using the HPC infrastructure LEO of the University of Innsbruck.

## Appendix
### A.1. Experimental Methods

He-tagging spectroscopy was carried out using the setup described elsewhere (Bergmeister et al. 2023). To create helium droplets around 100 nm in size, compressed and precooled helium was expanded through a 5 μm nozzle into a vacuum chamber. They were ionized by electron impact at 40 eV, generating multiply charged droplets (Laimer et al. 2019). Subsequently, the charged droplets were directed to the pick-up chamber, where a modified atomic carbon source was installed (Krasnokutski & Huisken 2014). The modification involved creating small openings in the tantalum filament. As a result, both $C_3$ and C species were simultaneously picked up by the He-droplets. Charge transfer from helium results in the formation of $C^+$ and $C_3^+$, while some portion of C and $C_3$ species may remain neutral. The following encounters of these species inside the droplets result in the growth of $C_n^+$ clusters in the same way as it happens in the ISM at low temperatures. In the next step, the doped He droplets arrived at the hexapole ion guide filled with helium gas. Helium droplets are heated and evaporated due to the collisions with He atoms, finally resulting in the formation of pure dopant ions and He-tagged dopant ions. The amount of He atoms attached to the dopant ions can be precisely controlled by adjusting the pressure of the He gas in the ion guide. The resulting mass spectra can be seen in Figure A1.

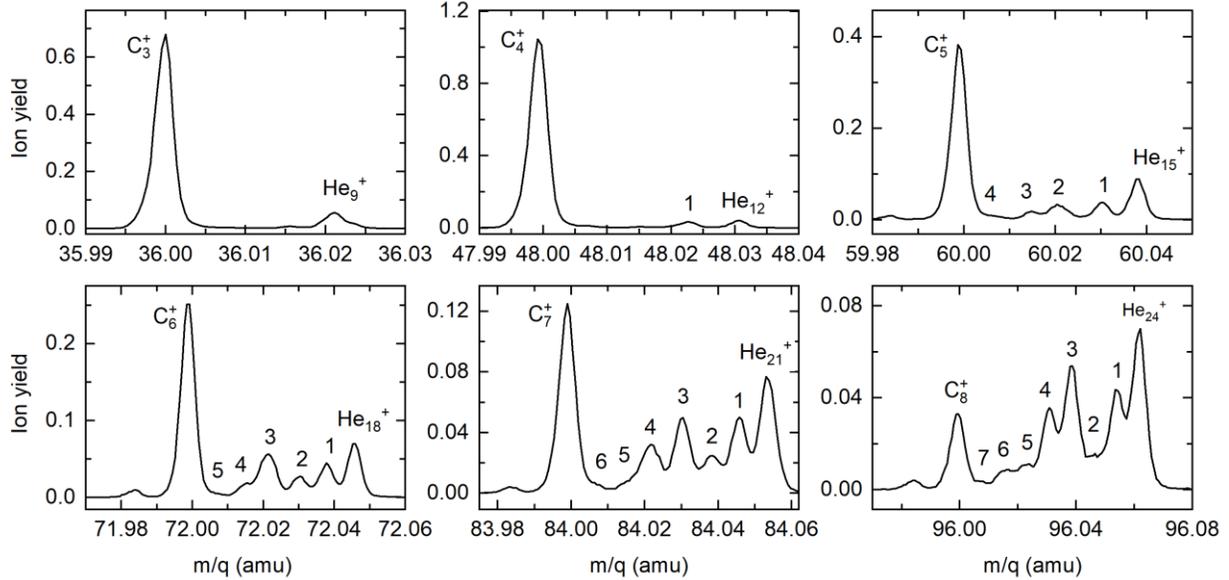

Figure A1. Mass spectra of He droplets with $C_3$ and C species around masses of $C_n$ clusters. Peaks corresponding to $C_n^+ He_m$ clusters are labeled with *n* numbers. The ion yields are normalized to the $C_4^+$ peak.

In the next chamber, a quadrupole mass filter (QMF) can be used to select ions with any specific mass range. These mass selected ions are allowed to interact with the light of a pulsed tunable laser (10kHz EKSPLA NT262) the beam of which is aligned parallel to the ion beam and overlaps it on about 35 cm. The laser was calibrated using a wavelength meter (SHR High-Resolution Wide-Range Spectrometer). All spectral data provided in the article are given in air wavelengths. The linewidth of the laser is less than 3 cm$^{-1}$ for wavelengths > 480nm and less than 5 cm$^{-1}$ for laser wavelengths < 480nm.
Finally, ions are analyzed by a high-resolution (m/Δm = 15000) time of flight (TOF) mass spectrometer. For the overview scans the QMF was set to pass all ions with the masses above the mass of $C_7^+$. This results in the presence of $C_4^+ He_m$ clusters with m from 11 to about 141. The spectroscopy was performed by tuning the wavelength of the laser and monitoring ion signals on the masses of

$C_n^+$. To obtain the position of the absorption bands of $C_4^+$ present in the gas phase, the QMF was set to let only $C_4^+He_m$ with specific $m$ ranges (m = 2 – 4, 2 – 12, 12 – 20, 26 – 34, m > 30) to pass. The spectroscopy is performed in the same way as before for overview scans by monitoring the appearance of $C_4^+$. For a small $m$, the shift of the absorption band depends linearly on $m$ (Kaiser, et al. 2018; Kappe, et al. 2023; Meyer, et al. 2021). Therefore, extrapolation to $m$ = 0 provides the position of the absorption bands in the gas phase. To correct the spectra for the variation of the laser power, we used a linear dependence between laser power and ion signal. This approach is validated since only a very small portion of He-tagged ions are photodissociated.

A.2. Theoretical Methods

Quantum chemical calculations of geometries and energies of different isomers of $C_4^+$ were performed using the CCSD method with the aug-cc-pVTZ correlation

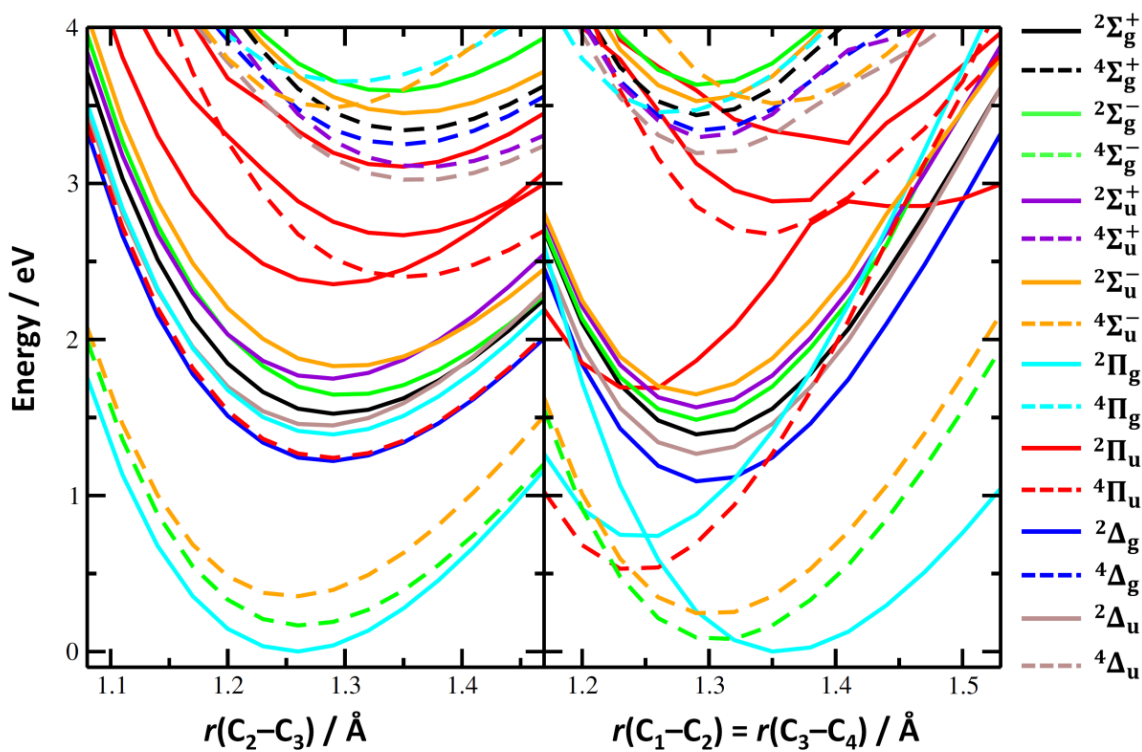

Figure A2. Calculated energies of electronic states for linear $C_4^+$ at the MRCI(9,9)/aug-cc-pVTZ level for the variation of the inner C-C bond (left) and the variation of the outer C-C bonds (right). Within the scans, the un-scanned middle and outer bonds were kept at 1.26 Å and 1.35 Å, respectively.

consistent basis set and B3LYP hybrid functional with the 6-311+G (d, p) basis set implemented in the GAUSSIAN16 package (Frisch et al. Gaussian Inc., 2016). The relative energies of isomers were obtained as differences between the sum of electronic and vibrational zero-point energies. The geometries of the ion in excited states were obtained using both the TDDFT-B3LYP method with the 6-311+G (d, p) basis set and the UEOMCCSD-FC method with the aug-cc-pVTZ basis set. To simulate the absorption spectrum, multidimensional Frank Condon factors were calculated using Gaussian16. Spectral broadening was simulated by giving each line a Gaussian line shape with a linewidth of 50 cm$^{-1}$. Boltzmann distributions were used to simulate spectra at 10 K. However, the optimization of excited states of cyclic $C_4^+$ was found to be complicated and cannot be performed due to strong wave functions mixing in $C_4^+$, which is in line with previous results (Hochlaf, et al. 2007). Additionally, we used the Multi-Reference Configuration

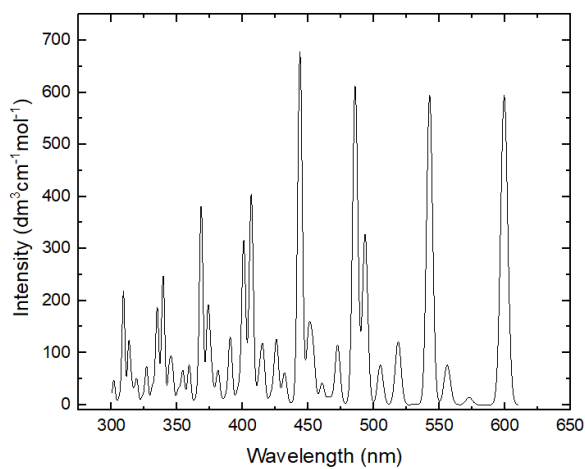

Figure A3. A calculated vibronic spectrum for $D_6 \leftarrow D_0(0)$ electronic transition of $C_4^+$ calculated at B3LYP/6-311G+(d,p) level. The 0-0 transition is set at position of 600 nm. It demonstrates a long vibrational progression, which is a characteristic of all linear $C_n^+$ chains.

Interaction Singles and Doubles (MRCI) to assess the position and shape of low-lying excited state potential energy surfaces (Figures A2, A4, Table A1). We used the active space of 9 electrons in 9 orbitals, denoted as (9,9), along with the aug-cc-pVTZ basis set. The multi-reference calculations were performed in Molpro (MOLPRO ; Werner et al. 2012; Werner et al. 2020).

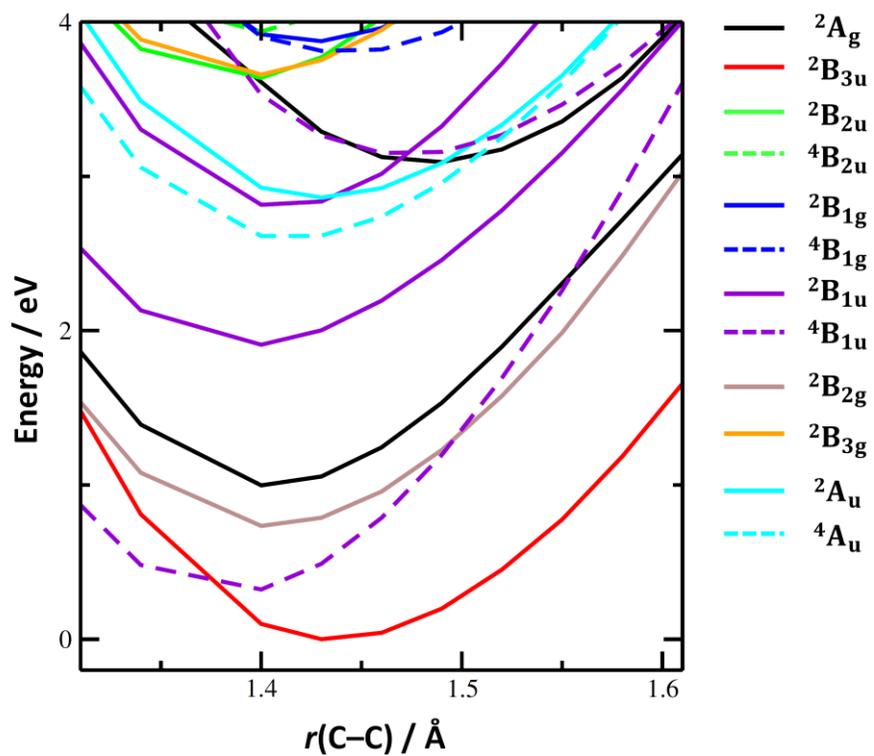

Figure A4. Calculated energies of electronic states for cyclic $C_4^+$ at the MRCI(9,9)/aug-cc-pVTZ level as a function of C-C distance. All C-C distances were kept equal, the inner C-C-C angles were kept as 111.4 and 68.6 degrees along the scan.

Table A1. Calculated energies (in eV) of electronic states in linear and cyclic $C_4^+$ ions as calculated at the MRCI(9,9)/aug-cc-pVTZ level of theory. The wavelengths of corresponding electronic transitions from the ground state are also provided. Linear isomer: C-C bond lengths of 1.35, 1.26, 1.35 Å. Cyclic isomer: C-C bond lengths of 1.43 Å, C-C-C angles of 111.4 and 68.6 degrees.

| $C_4^+$, linear | | | $C_4^+$, cyclic | | |
|---|---|---|---|---|---|
| | eV | nm | | eV | nm |
| $^2\Pi_g$ | 0.00 | | $^2B_{3u}$ | 0.00 | |
| $^4\Sigma_g^-$ | 0.17 | 7293 | $^4B_{1u}$ | 0.49 | 1569 |
| $^4\Sigma_u^-$ | 0.35 | 3542 | $^2B_{2g}$ | 0.79 | 1181 |
| $^2\Delta_g$ | 1.24 | 1000 | $^2A_g$ | 1.05 | 620 |
| $^4\Pi_u$ | 1.27 | 976 | $^2B_{1u}$ | 2.00 | 475 |
| $^2\Pi_g$ | 1.41 | 879 | $^4A_u$ | 2.61 | 437 |
| $^2\Delta_u$ | 1.46 | 849 | $^2B_{1u}$ | 2.84 | 434 |
| $^2\Sigma_g^+$ | 1.56 | 795 | $^2A_u$ | 2.86 | 380 |
| $^2\Sigma_g^-$ | 1.70 | 729 | $^4B_{1u}$ | 3.26 | 377 |
| $^2\Sigma_u^+$ | 1.77 | 700 | $^2A_g$ | 3.29 | 331 |
| $^2\Sigma_u^-$ | 1.88 | 659 | $^2B_{3g}$ | 3.75 | 329 |
| $^2\Pi_u$ | 2.39 | 519 | $^2B_{2u}$ | 3.77 | 325 |
| $^4\Pi_u$ | 2.67 | 464 | $^4B_{1g}$ | 3.81 | 320 |
| $^2\Pi_u$ | 2.89 | 429 | $^2B_{1g}$ | 3.88 | 308 |
| $^4\Delta_u$ | 3.31 | 375 | $^4B_{1g}$ | 4.03 | 306 |
| $^2\Pi_u$ | 3.34 | 371 | | | |
| $^4\Sigma_u^+$ | 3.44 | 360 | | | |
| $^4\Delta_g$ | 3.48 | 356 | | | |
| $^4\Sigma_u^-$ | 3.51 | 353 | | | |
| $^4\Sigma_g^+$ | 3.61 | 343 | | | |
| $^2\Sigma_u^-$ | 3.69 | 336 | | | |
| $^4\Pi_g$ | 3.70 | 335 | | | |
| $^2\Sigma_g^-$ | 3.77 | 329 | | | |